\documentclass[12pt]{article}

\usepackage{epsfig}

\begin{document}
\vspace*{-.6in}
\thispagestyle{empty}
\begin{flushright}
IASSNS-HEP-98/27 \\
CALT-68-2168
\end{flushright}
\baselineskip = 20pt

\vspace{.5in}
{\Large
\begin{center}
String Theory, Supersymmetry, Unification, and All That\footnote{To
appear in the American Physical Society Centenary issue of Reviews of
Modern Physics, March 1999}
\end{center}}
\medskip
\begin{center}
John H. Schwarz\\
\emph{California Institute of Technology, Pasadena, CA  91125 USA}\\
{\tt jhs@theory.caltech.edu}\\
and\\
Nathan Seiberg\\
\emph{Institute for Advanced Study, Princeton, NJ 08540 USA}\\
{\tt seiberg@ias.edu}
\end{center}
\vspace{0.5in}

\begin{center}
\textbf{Abstract}
\end{center}
\begin{quotation}
String theory and supersymmetry are theoretical ideas that go beyond
the standard model of particle physics and show promise for unifying
all forces.  After a brief introduction to supersymmetry, we discuss
the prospects for its experimental discovery in the near future.  We
then show how the magic of supersymmetry allows us to solve certain
quantum field theories exactly, thus leading to new insights about
field theory dynamics related to electric-magnetic duality.  The
discussion of superstring theory starts with its perturbation
expansion, which exhibits new features including ``stringy geometry.''
We then turn to more recent non-perturbative developments.  Using new
dualities, all known superstring theories are unified, and their
strong coupling behavior is clarified.  A central ingredient is the
existence of extended objects called branes.
\end{quotation}
\vfil
\newpage

\section{Introduction}

The standard model of particle physics (see the article by Gaillard,
Grannis, and Sciulli in this volume) is a beautiful theory that
accounts for all known phenomena up to energies of order 100~GeV.  Its
consistency relies on the intricacies of quantum field theory (see
Wilczek's article), and its agreement with experiment is spectacular.
However, there are many open problems with the standard model.  In
particular, we would like to know what lies beyond the standard model.
What is the physics at energies above 100~GeV?

One suggestion for physics at nearby energies of order 1~TeV (=
1000~GeV), which we will review below, is supersymmetry.  At higher
energies the various interactions of the standard model can be unified
into a grand unified theory (GUT).  Finally, at energies of order the
Planck energy $M_{P} c^2 =(c \hbar / G)^{1/2} c^2 \sim 10^{19}$~GeV
the theory must be modified.  This energy scale is determined on
dimensional grounds using Newton's constant $G$, the speed of light
$c$, and Planck's constant $\hbar$.  It determines the characteristic
energy scale of any theory that incorporates gravitation in a
relativistic and quantum mechanical setting.  At this energy scale the
gravitational interactions become strong and cannot be neglected.  How
to combine the elaborate structure of quantum field theory and the
standard model with Einstein's theory of gravity -- general relativity
-- is one of the biggest challenges in theoretical physics today.
String theory is the only viable attempt to achieve this!

There are various problems that arise when one attempts to combine
general relativity and quantum field theory.  The field theorist would
point to the breakdown of renormalizability -- the fact that
short-distance singularities become so severe that the usual methods
for dealing with them no longer work.  By replacing point-like
particles with one-dimensional extended strings, as the fundamental
objects, {\it superstring theory} certainly overcomes the problem of
perturbative non-renormalizability.  A relativist might point to a
different set of problems including the issue of how to understand the
causal structure of space-time when the metric has quantum-mechanical
fluctuations.  There are also a host of problems associated to black
holes such as the fundamental origin of their thermodynamic properties
and an apparent loss of quantum coherence.  The latter, if true, would
imply a breakdown in the basic structure of quantum mechanics.  The
relativist's set of issues cannot be addressed properly in a
perturbative setup, but recent discoveries are leading to
non-perturbative understandings that should help in addressing them.
Most string theorists expect that the theory will provide satisfying
resolutions of these problems without any revision in the basic
structure of quantum mechanics.  Indeed, there are indications that
someday quantum mechanics will be viewed as an implication of (or at
least a necessary ingredient of) superstring theory.

String theory arose in the late 1960's in an attempt to describe
strong nuclear forces.  In 1971 it was discovered that the inclusion
of fermions requires world-sheet supersymmetry.  This led to the
development of space-time supersymmetry, which was eventually
recognized to be a generic feature of consistent string theories --
hence the name {\it superstrings}.  String theory was a quite active
subject for about five years, but it encountered serious theoretical
difficulties in describing the strong nuclear forces, and QCD came
along as a convincing theory of the strong interaction.  As a result
the subject went into decline and was abandoned by all but a few
diehards for over a decade. In 1974 two of the diehards (Jo\"el Scherk
and JHS) proposed that the problems of string theory could be turned
into virtues if it were used as a framework for realizing Einstein's
old dream of {\it unification}, rather than as a theory of hadrons and
strong nuclear forces. In particular, the massless spin two particle in
the string spectrum, which had no sensible hadronic interpretation,
was identified as the graviton and shown to interact at low energies
precisely as required by general relativity.
One implication of this change in viewpoint was
that the characteristic size of a string became the Planck length,
$L_{P} = \hbar/ cM_P = (\hbar G/c^3)^{1/2} \sim 10^{-33} $~cm, some 20
orders of magnitude smaller than previously envisaged.  More refined
analyses lead to a string scale, $L_{S}$, that is a couple orders of
magnitude larger than the Planck length.  In any case, experiments at
existing accelerators cannot resolve distances shorter than about
$10^{-16}$~cm, which explains why the point-particle approximation of
ordinary quantum field theories is so successful.

\section{Supersymmetry}

Supersymmetry is a symmetry relating bosons and fermions according to
which every fermion has a bosonic superpartner and {\it vice versa}.
For example, fermionic quarks are partners of bosonic {\it squarks}.
By this we mean that quarks and squarks belong to the same irreducible
representation of the supersymmetry. Similarly, bosonic gluons (the
gauge fields of QCD) are partners of fermionic {\it gluinos}.  If
supersymmetry were an unbroken symmetry, particles and their
superpartners would have exactly the same mass. Since this is
certainly not the case, supersymmetry must be a broken symmetry (if it
is relevant at all).  In supersymmetric theories containing gravity,
such as supergravity and superstring theories, supersymmetry is a
gauge symmetry. Specifically, the superpartner of the graviton, called
the {\it gravitino}, is the gauge particle for local supersymmetry.

\subsection{Fermionic Dimensions of Spacetime}

Another presentation of supersymmetry is based on the notion of {\it
superspace}.  We do not change the structure of space-time but we add
structure to it.  We start with the usual four coordinates $X ^{\mu}=
t, x, y, z$ and add four odd dimensions $\theta_{\alpha}$ $ (\alpha =
1, \cdots, 4)$.  These odd dimensions are fermionic and anticommute
$$\theta_{\alpha}\theta_{\beta}= -\theta_{\beta}\theta_{\alpha}.$$
They are quantum dimensions that have no classical analog, which makes
it difficult to visualize or to understand them intuitively.  However,
they can be treated formally.

The fact that the odd directions are anticommuting has important
consequences.  Consider a function of superspace
$$\Phi (X, \theta) = \phi (X) + \theta_{\alpha} \psi_{\alpha}(X) +
\cdots + \theta^{4}F(X).$$ Since the square of any $\theta$ is zero
and there are only four different $\theta$'s, the expansion in powers
of $\theta$ terminates at the fourth order.  Therefore, a function of
superspace includes only a finite number of functions of $X$ (16 in
this case).  Hence, we can replace any function of superspace $\Phi
(X, \theta)$ with the component functions $\phi (X), \psi(X), \cdots$.
These include bosons $\phi(X), \cdots$ and fermions $\psi(X), \cdots$.
This is one way of understanding the pairing between bosons and
fermions.

A supersymmetric theory looks like an ordinary theory with degrees of
freedom and interactions that satisfy certain symmetry requirements.
Indeed, a supersymmetric quantum field theory is a special case of a
more generic quantum field theory rather than being a totally
different kind of theory. In this sense, supersymmetry by itself is
not a very radical proposal.  However, the fact that bosons and
fermions come in pairs in supersymmetric theories has important
consequences.  In some loop diagrams, like those in Fig.~1, the bosons
and the fermions cancel each other.  This boson-fermion cancellation
is at the heart of most of the applications of supersymmetry.  If
superpartners are present in the TeV range, this cancellation solves
the gauge hierarchy problem (see below).  Also, this cancellation is
one of the underlying reasons for being able to analyze supersymmetric
theories exactly.

\begin{figure}
\centerline{\psfig{figure=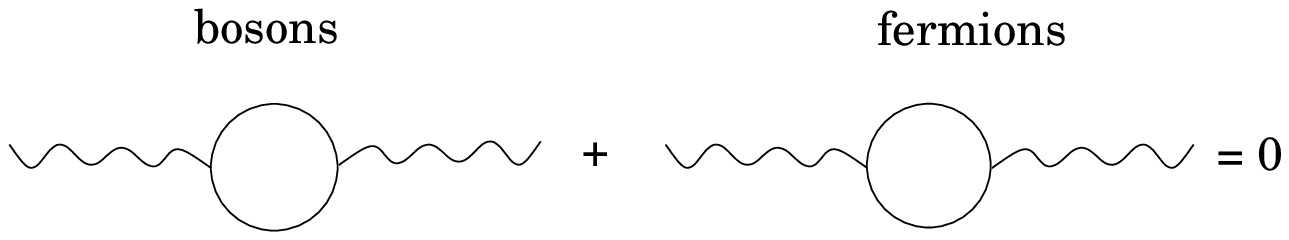,width=1in,angle=270}}
\caption[]{Boson-fermion cancellation in some loop diagrams}
\label{fig2}
\end{figure}

\subsection{Supersymmetry in the TeV Range}

There are several indications (discussed below) that supersymmetry is
realized in the TeV range so that the superpartners of the particles
of the standard model have masses of the order of a few TeV or
less. This is an important prediction, because the next generation of
experiments at Fermilab and CERN will explore the energy range where
at least some of the superpartners are expected to be found.
Therefore, within a decade or two we should know whether supersymmetry
exists at this energy scale.  If supersymmetry is indeed discovered in
the TeV range, this will amount to the discovery of the new odd
dimensions.  This will be a major change in our view of space and
time.  It would be a remarkable success for theoretical physics --
predicting such a deep notion without any experimental input!

\subsubsection{The Gauge Hierarchy Problem}

The {\it gauge hierarchy problem} is essentially a problem of
dimensional analysis.  Why is the characteristic energy of the
standard model, which is given by the mass of the W boson $M_{W} \sim
100$~GeV, so much smaller than the the characteristic scale of
gravity, the Planck mass $M_{P} \sim 10^{19}$~GeV?  It should be
stressed that in quantum field theory this problem is not merely an
aesthetic problem, but also a serious technical problem.  Even if such
a hierarchy is present in some approximation, radiative corrections
tend to destroy it.  More explicitly, divergent loop diagrams restore
dimensional analysis and move $M_{W} \rightarrow M_{P}$.

The main theoretical motivation for supersymmetry at the TeV scale is
the hierarchy problem.  As we mentioned, in supersymmetric theories
some loop diagrams vanish -- or become less divergent -- due to
cancellations between bosons and fermions.  In particular the loop
diagram restoring dimensional analysis is cancelled as in Fig.~1.
Therefore, in its simplest form supersymmetry solves the technical
aspects of the hierarchy problem.  More sophisticated ideas, known as
dynamical supersymmetry breaking, also solve the aesthetic problem.

\subsubsection{The Supersymmetric Standard Model}

The minimal supersymmetric extension of the standard model (the MSSM)
contains superpartners for all the particles of the standard model, as
we have already indicated.  Some of their coupling constants are
determined by supersymmetry and the known coupling constants of the
standard model.  Most of the remaining coupling constants and the
masses of the superpartners depend on the details of supersymmetry
breaking.  These parameters are known as {\it soft breaking terms}.
Various phenomenological considerations already put strong constraints
on these unknown parameters but there is still a lot of freedom in
them.  If supersymmetry is discovered, the new parameters will be
measured.  These numbers will be extremely interesting as they will
give us a window into physics at higher energies.

The MSSM must contain two electroweak doublets of Higgs
fields. Whereas a single doublet can give mass to all quarks and
charged leptons in the standard model, the MSSM requires one doublet
to give mass to the charge 2/3 quarks and another to give mass to the
charge -1/3 quarks and charged leptons. Correspondingly, electroweak
symmetry breaking by the Higgs mechanism involves two Higgs fields
obtaining vacuum expectation values. The ratio, called $\tan \beta$,
is an important phenomenological parameter.  In the standard model the
Higgs mass is determined by the Higgs vacuum expectation value and the
strength of Higgs self coupling (coefficient of the $\phi^4$ term in
the potential).  In supersymmetry the latter is related to the
strength of the gauge interactions.  This leads to a prediction for
the mass of the lightest Higgs boson $h$ in the MSSM. In the leading
semiclassical approximation one can show that $M_h \leq M_Z |\cos 2
\beta |$, where $M_Z \sim$ 91~GeV is the mass of the Z boson. Due to
the large mass of the top quark, radiative corrections to this bound
can be quite important.  A reasonably safe estimate is that $M_h \leq
130$~GeV, which should be compared to current experimental lower
bounds of about 80~GeV. The discovery of a relatively light Higgs
boson, which might precede the discovery of any superparticles, would
be encouraging for supersymmetry.  However, it should be pointed out
that there are rather mild extensions of the MSSM where the upper
bound is significantly higher.

It is useful to assign positive {\it R parity} to the known particles
(including the Higgs) of the standard model and negative R parity to
their superpartners.  For reasonable values of the new parameters
(including the soft breaking terms) R parity is a good symmetry.  In
this case the lightest supersymmetric particle (called the LSP) is
absolutely stable.  It could be an important constituent of the dark
matter of the Universe.

\subsubsection{Supersymmetric Grand Unification}

\begin{figure}
\centerline{\psfig{figure=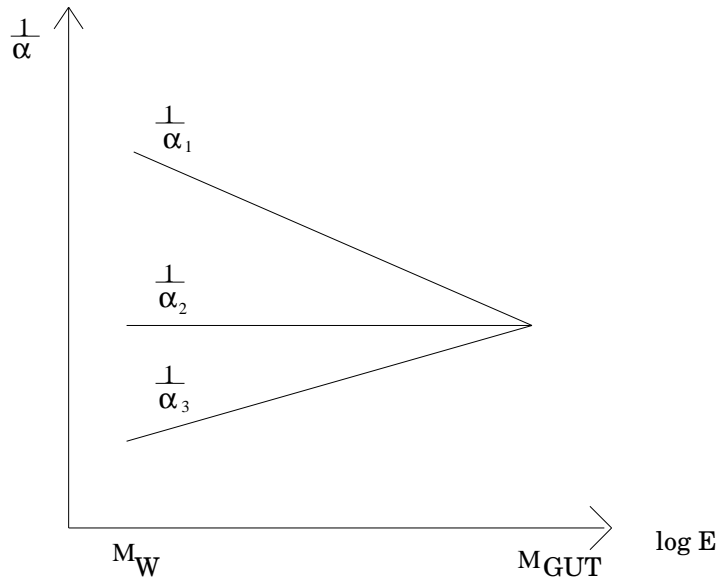,width=3in,angle=270}}
\caption[]{Coupling constant unification in supersymmetric theories}
\label{fig3}
\end{figure}

The second motivation for supersymmetry in the TeV range comes from
the idea of gauge unification.  Recent experiments have yielded
precise determinations of the strengths of the $SU(3)\times
SU(2)\times U(1)$ gauge interactions -- the analogs of the fine
structure constant for these interactions.  They are usually denoted
by $\alpha_3$, $\alpha_2$ and $\alpha_1$ for the three factors in
$SU(3)\times SU(2)\times U(1)$.  In quantum field theory these values
depend on the energy at which they are measured in a way that depends
on the particle content of the theory.  Using the measured values of
the coupling constants and the particle content of the standard model,
one can extrapolate to higher energies and determine the coupling
constants there.  The result is that the three coupling constants do
not meet at the same point.  However, repeating this extrapolation
with the particles belonging to the minimal supersymmetric extension
of the standard model, the three gauge coupling constants meet at a
point, $M_{\rm GUT}$, as sketched in Fig.~2.  At that point the
strengths of the various gauge interactions become equal and the
interactions can be unified into a {\it grand unified theory}.
Possible grand unified theories embed the known $SU(3)\times
SU(2)\times U(1)$ gauge group into $SU(5)$ or $SO(10)$.

How much significance should we assign to this result?  Two lines must
meet at a point.  Therefore, there are only two surprises here.  The
first is that the third line intersects the same point.  The second
more qualitative one is that the unification scale, $M_{\rm GUT}$, is
at a reasonable value.  Its value is consistent with the experimental
bound from proton decay, and it is a couple of orders of magnitude
below the Planck scale, where gravity would need to be taken into
account.  One could imagine that that there are other modifications of
the standard model that achieve the same thing, so this is far from a
proof of supersymmetry, but it is certainly encouraging circumstantial
evidence. It is an independent indication that superpartner masses
should be around a TeV.

\subsection{Supersymmetric Quantum Field Theories}

Quantum field theory is notoriously complicated.  It is a non-linear
system of an infinite number of coupled degrees of freedom.
Therefore, until recently when the power of supersymmetry began to be
exploited, there were few exact results for quantum field theories
(except in two dimensions).  However, it has been realized recently
that a large class of physical quantities in many supersymmetric
quantum field theories can be computed exactly by analytic methods!

The main point is that these theories are very constrained.  The
dependence of some observables on the parameters of the problem is so
constrained that there is only one solution that satisfies all of the
consistency conditions.  More technically, because of supersymmetry
some observables vary holomorphically (complex analytically) with the
coupling constants, which are complex numbers in these theories.  Due
to Cauchy's theorem, such analytic functions are determined in terms
of very little data: the singularities and the asymptotic behavior.
Therefore, if supersymmetry requires an observable to depend
holomorphically on the parameters and we know the singularities and
the asymptotic behavior, we can determine the exact answer.  The
boson-fermion cancellation, which we mentioned above in the context of
the hierarchy problem, can also be understood as a consequence of a
constraint following from holomorphy.

\subsubsection{Families of Vacua}

Another property of many supersymmetric theories that makes them
tractable is that they have a family of inequivalent vacua.  To
understand this fact we should contrast it with the situation in a
ferromagnet, which has a continuum of vacua, labeled by the common
orientation of the spins.  These vacua are all equivalent; {\it i.e.,}
the physical observables in one of these vacua are exactly the same as
in any other.  The reason is that these vacua are related by a
symmetry. The system must choose one of them, which leads to
spontaneous symmetry breaking.

\begin{figure}
\centerline{\psfig{figure=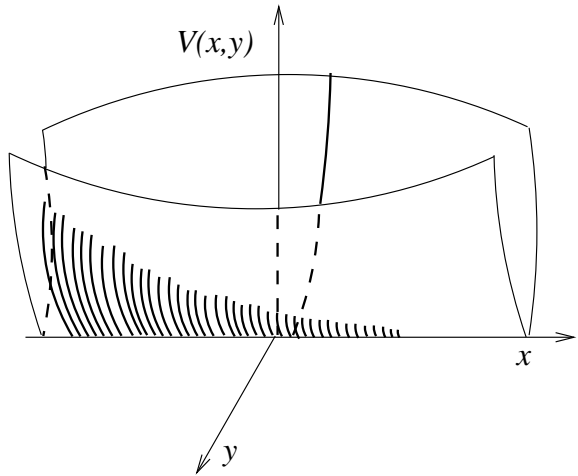,width=3in}}
\caption[]{Typical potential in supersymmetric theories exhibiting
``accidental vacuum degeneracy''}
\label{fig4}
\end{figure}

We now study a situation with inequivalent vacua in contrast to the
ferromagnet.  Consider the case in which degrees of freedom, called
$x$ and $y$, have the potential $V(x,y)$ shown in Fig.~3.  The vacua
of the system correspond to the different points along the valley of
the potential, $y=0$ with arbitrary $x$.  However, as we tried to make
clear in the figure, these points are inequivalent -- there is no
symmetry that relates them.  More explicitly, the potential is shallow
around the origin but becomes steep for large $x$.  Such {\it
accidental degeneracy} is usually lifted by quantum effects.  For
example, if the system corresponding to the potential in the figure
has no fermions, the zero-point fluctuations around the different
vacua would be different.  They would lead to a potential along the
valley pushing the minimum to the origin.  However, in a
supersymmetric theory the zero-point energy of the fermions exactly
cancels that of the bosons and the degeneracy is not lifted.  The
valleys persist in the full quantum theory.  Again, we see the power
of the boson-fermion cancellation.  We see that a supersymmetric
system typically has a continuous family of vacua.  This family, or
manifold, is referred to as a {\it moduli space of vacua}, and the
modes of the system corresponding to motion along the valleys are
called {\it moduli}.

The analysis of supersymmetric theories is usually simplified by the
presence of these manifolds of vacua.  Asymptotically, far along the
flat directions of the potential, the analysis of the system is simple
and various approximation techniques are applicable.  Then, by using
the asymptotic behavior along several such flat directions, as well as
the constraints from holomorphy, a unique solution is obtained.  This
is a rather unusual situation in physics.  We perform approximate
calculations, which are valid only in some regime, and this gives us
the exact answer.  This is a theorist's heaven -- exact results with
approximate methods!

\subsubsection{Electric-Magnetic Duality}

Once we know how to solve such theories, we can analyze many examples.
The main lesson that has been learned is the fundamental role played by
{\it electric-magnetic duality.}  It turns out to be the underlying
principle controlling the dynamics of these systems.

When faced with a complicated system with many coupled degrees of
freedom it is common in physics to look for weakly coupled variables
that capture most of the phenomena.  For example, in condensed matter
physics we formulate the problem at short distance in terms of
interacting electrons and nuclei.  The desired solution is the
macroscopic behavior of the matter and its possible phases.  It is
described by weakly coupled effective degrees of freedom.  Usually
they are related in a complicated, and in most cases unknown, way to
the microscopic variables.  Another example is hydrodynamics, where
the microscopic degrees of freedom are molecules and the long distance
variables are properties of a fluid that are described by partial
differential equations.

In one class of supersymmetric field theories the long distance
behavior is described by a set of weakly coupled effective degrees of
freedom.  These are composites of the elementary degrees of freedom.
As the characteristic length scale becomes longer, the interactions
between these effective degrees of freedom become weaker, and the
description in terms of them becomes more accurate.  In other words,
the long distance theory is a ``trivial'' theory in terms of the
composite effective degrees of freedom.

In another class of examples there are no variables in terms of which
the long distance theory is simple -- the theory remains interacting.
Because it is scale invariant, it is at a non-trivial fixed point of
the renormalization group.  In these situations there are two (or
more) dual descriptions of the physics leading to identical results
for the long distance interacting behavior.

In both classes of examples an explicit relation between the two sets
of variables is not known.  However, there are several reasons to
consider these pairs of descriptions as being electric-magnetic duals
of one another.  The original variables at short distance are referred
to as the {\it electric} degrees of freedom and the other set of
long-distance variables as the {\it magnetic} ones.  These two dual
descriptions of the same theory give us a way to address strong
coupling problems.  When the electric variables are strongly coupled,
they fluctuate rapidly and their dynamics is complicated (see the
table below).  However, then the magnetic degrees of freedom are
weakly coupled.  They do not fluctuate rapidly and their dynamics is
simple.  In the first class of examples the magnetic degrees of
freedom are the macroscopic ones which are free at long distance.
They are massless bound states of the elementary particles.  In the
second class of examples there are two valid descriptions of the long
distance theory: electric and magnetic.  Since both of them are
interacting, neither of them gives a ``trivial'' description of the
physics.  However, as one of them becomes more strongly coupled, the
other becomes more weakly coupled (see the table below).

Finally, using this electric-magnetic duality we can find a simple
description of complicated phenomena associated with the phase diagram
of the theories.  For example, as the electric degrees of freedom
become strongly coupled, they can lead to confinement.  In the
magnetic variables, this is simply the Higgs phenomenon
(superconductivity) which is easily understood in weak coupling.
The electric-magnetic relations are summarized in the following table:
\begin{table}[hb]
$$\vbox {\rm \halign {\strut#&\vrule#&\quad\hfil #\hfil\quad&\vrule#& 
\quad\hfil #\hfil\quad &\vrule#&\quad\hfil #\hfil\quad&\vrule#
\cr\noalign{\hrule}
&&& &electric && magnetic & \cr\noalign{\hrule}
&&coupling & &strong & &weak & \cr\noalign{\hrule}
&&fluctuations && large && small & \cr\noalign{\hrule}
&&phase & &confinement && Higgs & \cr\noalign{\hrule}
}}$$
\end{table}

Apart from the ``practical'' application to solving quantum field
theories, the fact that a theory can be described either in terms of
electric or magnetic variables has deep consequences:
\begin{itemize}
\item
For theories belonging to the first class of examples it is natural to
describe the magnetic degrees of freedom as composites of the
elementary electric ones.  The magnetic particles typically include
massless gauge particles reflecting a new magnetic gauge symmetry.
These massless composite gauge particles are associated with a new
gauge symmetry which is not present in the fundamental electric
theory.  Since this gauge symmetry is not a symmetry of the original,
short distance theory, it is generated by the dynamics rather than
being ``put in by hand.''  We see that, in this sense, 
{\it gauge invariance cannot be fundamental.}

\item
For theories of the second class the notion of elementary particle
breaks down.  {\it There is no invariant way of choosing which degrees
of freedom are elementary and which are composite.} The magnetic
degrees of freedom can be regarded as composites of the electric ones
and {\it vice versa}. \end{itemize}

\section{Superstring Theory}

\subsection{Perturbative String Theory}

All superstring theories contain a massless scalar field, called the
{\it dilaton} $\phi$, that belongs to the same supersymmetry multiplet
as the graviton. In the semiclassical approximation, this field
defines a flat direction in the moduli space of vacua, so that it can
take any value $\phi_0$.  Remarkably, this determines the string
coupling constant $g_S = e^{\phi_0}$, which is a dimensionless
parameter on which one can base a perturbation expansion.  The
perturbation expansions are power series expansions in powers of the
string coupling constant like those that are customarily used to carry
out computations in quantum field theory.

\subsubsection{Structure of the String World-Sheet and the
Perturbation Expansion}

\begin{figure}
\centerline{\psfig{figure=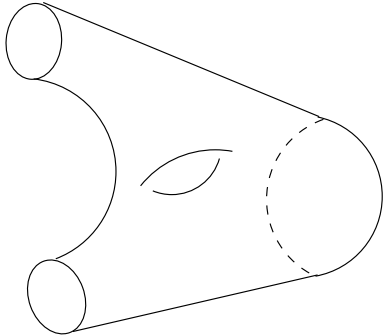,width=2in,angle=180}}
\caption[]{An example of a string world-sheet with two initial
strings, one final string and a handle}
\label{ws}
\end{figure}

A string's space-time history is described by functions $X^\mu
(\sigma,\tau)$ that map the string's two-dimensional {\it world sheet}
$(\sigma,\tau)$ into space-time $X^\mu$.  There are also other
world-sheet fields that describe other degrees of freedom, such as
those associated with supersymmetry and gauge symmetries.
Surprisingly, {\em classical} string theory dynamics is described by a
conformally invariant 2d {\em quantum} field theory.  What
distinguishes one-dimensional strings from higher-dimensional analogs
(discussed later) is the fact that this 2d theory is renormalizable.
Perturbative quantum string theory can be formulated by the Feynman
sum-over-histories method.  This amounts to associating a genus $h$
Riemann surface (a closed and orientable two-dimensional surface with
$h$ handles) to an $h$-loop string theory Feynman diagram. It contains
a factor of $g_S^{2h}$.  For example, the string world sheet in Fig.~4
has one handle.

The attractive features of this approach are that there is just one
diagram at each order $h$ of the perturbation expansion and that each
diagram represents an elegant (though complicated) finite-dimensional
integral that is ultraviolet finite.  In other words, they do not give
rise to the severe short distance singularities that plague other
attempts to incorporate general relativity in a quantum field theory.
The main drawback of this approach is that it gives no insight into
how to go beyond perturbation theory.

\subsubsection{Five Superstring Theories}

In 1984-85 there was a series of discoveries that convinced many
theorists that superstring theory is a very promising approach to
unification.  This period is now sometimes referred to as {\it the
first superstring revolution}.  Almost overnight, the subject was
transformed from an intellectual backwater to one of the most active
areas of theoretical physics, which it has remained ever since.  By
the time the dust settled, it was clear that there are five different
superstring theories, each requiring ten dimensions (nine space and
one time), and that each has a consistent perturbation expansion.  The
five theories are denoted type I, type IIA, type IIB, $E_8 \times E_8$
heterotic (HE, for short), and SO(32) heterotic (HO, for short).  The
type II theories have two supersymmetries in the ten-dimensional
sense, while the other three have just one.  The type I theory is
special in that it is based on unoriented open and closed strings,
whereas the other four are based on oriented closed strings. Type I
strings can break, whereas the other four are unbreakable.  The IIA
theory is non-chiral ({\it i.e.}, it is parity conserving), and the
other four are chiral (parity violating).

\subsubsection{Compactification of Extra Dimensions}

To have a chance of being realistic, the six extra space dimensions
must somehow curl up into a tiny geometrical space as in {\it
Kaluza--Klein theory}.  The linear size of this space is presumably
comparable to the string scale $L_{S}$.  Since space-time geometry is
determined dynamically (as in general relativity) only geometries that
satisfy the dynamical equations are allowed. Among such solutions, one
class stands out: The HE string theory, compactified on a particular
kind of six-dimensional space, called a {\it Calabi--Yau manifold},
has many qualitative features at low energies that resemble the
supersymmetric extension of the standard model of elementary
particles.  In particular, the low mass fermions occur in suitable
representations of a plausible unifying gauge group. Moreover, they
occur in families whose number is controlled by the topology of the
Calabi--Yau manifold.  These successes have been achieved in a
perturbative framework, and are necessarily qualitative at best, since
non-perturbative phenomena are essential to an understanding of
supersymmetry breaking and other important details.

\subsubsection{T Duality and Stringy Geometry}

The basic idea of {\it T duality} can be illustrated by considering a
compact spatial dimension consisting of a circle of radius $R$.  In
this case there are two kinds of excitations to consider.  The first,
which is not special to string theory, is due to the quantization of
the momentum along the circle.  These {\it Kaluza--Klein} excitations
contribute $(n/R)^2$ to the energy squared, where $n$ is an integer.
The second kind are {\it winding-mode} excitations, which arise due to
a closed string being wound $m$ times around the circular dimension.
They are special to string theory, though there are higher-dimensional
analogs.  Letting $T = (2\pi L_{S}^2)^{-1}$ denote the fundamental
string {\it tension} (energy per unit length), the contribution of a
winding mode to the energy squared is $(2\pi R m T)^2$.  T duality
exchanges these two kinds of excitations by mapping $m \leftrightarrow
n$ and $R \leftrightarrow L_{S}^2 /R$.  This is part of an exact map
between a T-dual pair of theories A and B.

We see that the underlying geometry is ambiguous -- there is no way to
tell the difference between a compactification on a circle of radius
$R$ and a compactification on a circle of radius $ L_{S}^2 /R$.  This
ambiguity is clearly related to the fact that the objects used to
probe the circle are extended objects -- strings -- which can wind
around the circle.

One implication of this ambiguity is that usual geometric concepts
break down at short distances, and classical geometry is replaced by
{\it stringy geometry}, which is described mathematically by 2d
conformal field theory.  It also suggests a generalization of the
Heisenberg uncertainty principle according to which the best possible
spatial resolution $\Delta x$ is bounded below not only by the
reciprocal of the momentum spread, $\Delta p$, but also by the string
size, which grows with energy.  This is the best one can do using
fundamental strings as probes. However, by probing with certain
non-perturbative objects called {\it D-branes}, which we will discuss
later, it is sometimes possible (but not in the case of the circle
discussed above) to do better.

A closely related phenomenon is that of {\it mirror symmetry}.  In the
example of the circle above the topology was not changed by T duality.
Only the size was transformed.  In more complicated compactifications,
like compactifications on Calabi-Yau manifolds, there is even an
ambiguity in the underlying topology -- there is no way to tell on
which of two mirror pairs of Calabi-Yau manifolds the theory is
compactified.  This ambiguity can be useful because it is sometimes
easier to perform some calculations with one Calabi-Yau manifold than
with its mirror manifold.  Then, using mirror symmetry we can infer
what the answers are for different compactifications.

\begin{figure}[t]
\centerline{\epsfxsize=5truein \epsfbox{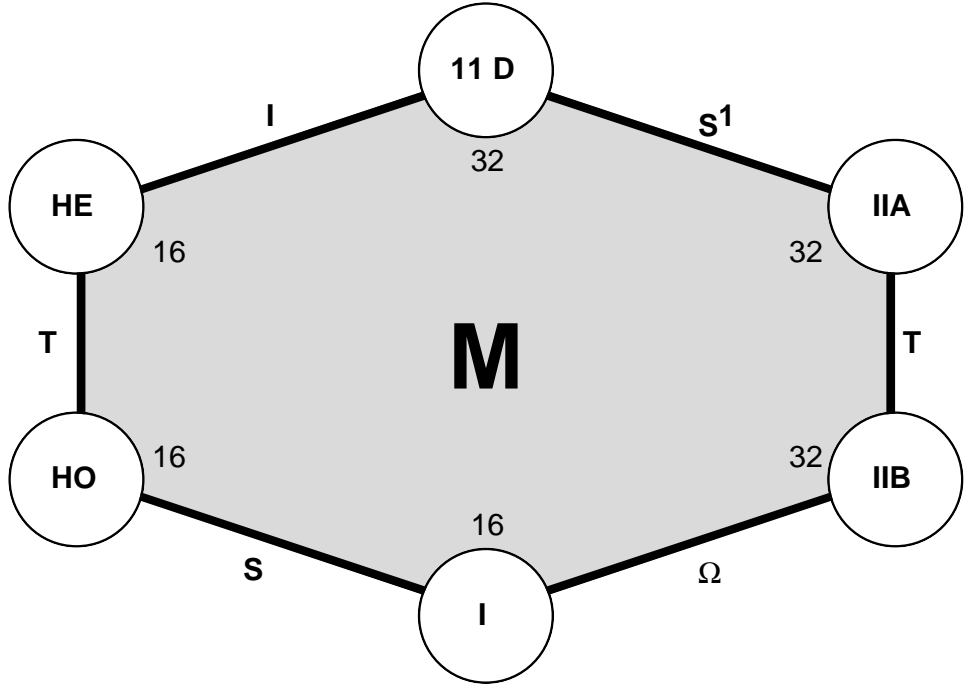} }
\caption{The M  theory moduli space.}
\end{figure}

Two pairs of ten-dimensional superstring theories are T dual when
compactified on a circle: the IIA and IIB theories and the HE and HO
theories.  The two edges of Fig.~5 labeled T connect vacua related by
T duality.  For example, if the IIA theory is compactified on a circle
of radius $R_A$ leaving nine noncompact dimensions, this is equivalent
to compactifying the IIB theory on a circle of radius $R_B =
L_S^2/R_A$. The T duality relating the two heterotic theories (HE and
HO) is essentially the same, though there are additional technical
details in this case.

Another relation between theories is the following.  A
compactification of the type I theory on a circle of radius $R_I$
turns out to be related to a certain compactification of the type IIA
on a line interval I with size proportional to $L_S^2/R_I$.  The line
interval can be thought of as a circle with some identification of
points ${\rm I} = S^1/\Omega$.  Therefore, we can say that the type I
theory on a circle of radius $R_I$ is obtained from the type IIA on a
circle of radius $L_S^2/R_I$ by acting with $\Omega$.  Since by T
duality the IIA theory on a circle of radius $L_S^2/R_I$ is the same
as the IIB theory on a circle of radius $R_I$, we conclude that upon
compactification on a circle type I is obtained from IIB by the action
of $\Omega$.  By taking $R_I$ to infinity this relation is also true
in 10 dimensions.  This is the reason for the edge denoted by $\Omega$
in Fig.~5.

These dualities reduce the number of (apparently) distinct superstring
theories from five to three, or if we also use $\Omega$ to two.  The
point is that the two members of each pair are continuously connected
by varying the compactification radius from zero to infinity. Like the
string coupling constant, the compactification radius arises as the
value of a scalar field.  Therefore varying this radius is a motion in
the moduli space of quantum vacua rather than a change in the
parameters of the theory.

\subsection{Non-Perturbative String Theory}

The {\it second superstring revolution} (1994-??) has brought
non-perturbative string physics within reach.  The key discoveries
were various {\it dualities}, which show that what was viewed
previously as five distinct superstring theories is in fact five
different perturbative expansions of a single underlying theory about
five different points in the moduli space of consistent vacua!  It is
now clear that there is a unique theory, though it allows many
different vacua.  A sixth special vacuum involves an 11-dimensional
Minkowski space-time. Another lesson we have learned is that,
non-perturbatively, objects of more than one dimension (membranes and
higher {\it p-branes}) play a central role.  In most respects they
appear to be on an equal footing with strings, but there is one big
exception: a perturbation expansion cannot be based on p-branes with
$p > 1$.

A schematic representation of the relationship between the five
superstring vacua in 10d and the 11d vacuum is given in Fig.~5.  The
idea is that there is some large moduli space of consistent vacua of a
single underlying theory -- denoted by M here.  The six limiting
points, represented as circles, are special in the sense that they are
the ones with (super) Poincar\'e invariance in ten or eleven
dimensions.  The letters on the edges refer to the type of duality
relating a pair of limiting points. The numbers 16 or 32 refer to the
number of unbroken supersymmetries.  In 10d the minimal spinor has 16
real components, so the conserved supersymmetry charges (or {\it
supercharges}) correspond to just one spinor in three cases (type I,
HE, and HO). Type II superstrings have two such spinorial
supercharges.  In 11d the minimal spinor has 32 real components.

\subsubsection{S Duality}

Suppose now that a pair of theories (A and B) are {\it S dual}. This
means that if $f_A(g_S)$ denotes any physical observable of theory A,
where $g_S$ is the coupling constant, then there is a corresponding
physical observable $f_B(g_S)$ in theory B such that $f_A (g_S) = f_B
(1/g_S)$.  This duality relates one theory at weak coupling to the
other at strong coupling. It generalizes the electric-magnetic duality
of certain field theories, discussed in section 2.  S duality relates
the type I theory to the HO theory and the IIB theory to itself.  This
determines the strong coupling behavior of these three theories in
terms of weakly coupled theories. Varying the strength of the string
coupling also corresponds to a motion in the moduli space of vacua.

The edge connecting the HO vacuum and the type I vacuum is labeled by
S in the diagram, since these two vacua are related by S duality.  It
had been known for a long time that the two theories have the same
gauge symmetry ($SO(32)$) and the same kind of supersymmetry, but it
was unclear how they could be equivalent, because type I strings and
heterotic strings are very different.  It is now understood that
$SO(32)$ heterotic strings appear as non-perturbative excitations in
the type I description.

\subsubsection{M Theory and the Eleventh Dimension}

The understanding of how the remaining two superstring theories (type
IIA and HE) behave at strong coupling came as quite a surprise.  In
each case there is an 11th dimension whose size $R$ becomes large at
strong string coupling $g_S$.  In the IIA case the 11th dimension is a
circle, whereas in the HE case it is a line interval.  The strong
coupling limit of either of these theories gives an 11-dimensional
Minkowski space-time.  The eleven-dimensional description of the
underlying theory is called {\it M theory}.\footnote{The letter M
could stand for a variety of things such as magic, mystery, meta,
mother, or membrane.}

The 11d vacuum, including 11d supergravity, is characterized by a
single scale -- the 11d Planck scale $L_P$.  It is proportional to
$G^{1/9}$, where $G$ is the 11d Newton constant.  The connection to
type IIA theory is obtained by taking one of the ten spatial
dimensions to be a circle ($S^1$ in the diagram) of radius $R$. As we
pointed out earlier, the type IIA string theory in 10d has a
dimensionless coupling constant $g_S$, given by the value of the
dilaton field, and a length scale, $L_S$.  The relationship between
the parameters of the 11d and IIA descriptions is given by
\begin{equation} \label{ems}
L_P^3 = R L_S^2
\end{equation}
\begin{equation} \label{gs}
R= L_S g_S.
\end{equation}
Numerical factors (such as $2\pi$) are not important for present
purposes and have been dropped.  The significance of these equations
will emerge later.  However, one point can be made immediately.  The
conventional perturbative analysis of the IIA theory is an expansion
in powers of $g_S$ with $L_S$ fixed.  The second relation implies that
this is an expansion about $R = 0$, which accounts for the fact that
the 11d interpretation was not evident in studies of perturbative
string theory.  The radius $R$ is a modulus -- the value of a massless
scalar field with a flat potential.  One gets from the IIA point to
the 11d point by continuing this value from zero to infinity.  This is
the meaning of the edge of Fig.~5 labeled $S^1$.

The relationship between the HE vacuum and 11d is very similar.  The
difference is that the compact spatial dimension is a line interval
(denoted I in the Fig.~5) instead of a circle.  The same relations in
eqs. (\ref{ems}) and (\ref{gs}) apply in this case.  This
compactification leads to an 11d space-time that is a slab with two
parallel 10d faces.  One set of $E_8$ gauge fields is confined to each
face, whereas the gravitational fields reside in the bulk.  One of the
important discoveries in the first superstring revolution was a
mechanism that cancels quantum mechanical anomalies in the Yang-Mills
and Lorentz gauge symmetries. This mechanism only works for $SO(32)$
and $E_8\times E_8$ gauge groups.  There is a nice generalization of
this 10d anomaly cancellation mechanism to the setting of 11
dimensions with a 10d boundary.  It only works for $E_8$ gauge groups!

\subsubsection{p-branes and D-branes}

In addition to the strings the theory turns out to contain other
objects, called {\it p-branes}.  A p-brane is an extended object in
space with p spatial dimensions.  (The term p-brane originates from
the word membrane which describes a 2-brane.)  For example, the 11d M
theory turns out to contain two basic kinds of p-branes with $p=2$ and
$p=5$, called the M2-brane and the M5-brane.  A simpler example of a
brane is readily understood in the type IIA theory when it is viewed
as a compactification of the 11d theory on a circle.
Eleven-dimensional particles with momentum around the circle appear as
massive particles in 10d, whose masses are proportional to $1/R$.
Since they are point particles, they are referred to as 0-branes.
Using eq.~(\ref{gs}), $1/R=1/L_Sg_S$, and we see that in the
perturbative string region, where $g_S \ll 1$, they are much heavier
than the ordinary string states whose masses are of order $1/L_S$.
The type IIA string in 10 dimensions can be identified as the M2-brane
wrapping the compact circle.

These p-branes are crucial in the various dualities discussed above --
since they are states in the theory, they should be mapped correctly
under T and S dualities.  This is particularly interesting for S
duality, which maps the fundamental string of one theory to a heavy
1-brane of the other.  For example, the heterotic string is such a
heavy 1-brane in the weakly coupled type I theory.  We therefore see
that the notion of an elementary (or fundamental) string is
ill-defined.  The string which appears fundamental at one boundary of
Fig.~5 is a heavy brane at another boundary and vice-versa.  We have
already encountered a similar phenomenon in our discussion of
electric-magnetic duality in field theory, where there was an
ambiguity in the notion of elementary objects.

A special class of
p-branes are called {\it Dirichlet p-branes} (or {\it D-branes} for
short).  The name derives from the boundary conditions assigned to the
ends of open strings.  The usual open strings of the type I theory
have Neumann boundary conditions at their ends.  More generally, in
type II theories, one can consider an open string with boundary
conditions at the end given by $\sigma = 0$
$$ {\partial X^\mu\over\partial\sigma} = 0  \quad\quad\mu =
0,1,\ldots, p $$
$$ X^\mu = X_0^\mu \quad\quad\mu = p + 1, \ldots, 9 $$ 
and similar boundary conditions at the other end.  The interpretation
of these equations is that strings end on a $p$-dimensional object in
space -- a D-brane. The description of D-branes as a place where open
strings can end leads to a simple picture of their dynamics.  For weak
string coupling this enables the use of perturbation theory to study
non-perturbative phenomena!

D-branes have found many interesting applications. One of the most
remarkable of these concerns the study of black holes.  Specifically,
D-brane techniques can be used to count the quantum microstates
associated to classical black hole configurations and to show that in
suitable limits the entropy (defined by $S = \log N$, where $N$ is the
number of quantum states the system can be in) agrees with the
Bekenstein--Hawking prediction: 1/4 the area of the event horizon.
For further details, see the article by Horowitz and Teukolsky.

D-branes also led to new insights and new results in quantum field
theory.  This arises from the realization that the open strings which
end on D-branes are described at low energies by a local quantum field
theory ``living'' on the brane.  The dynamics of quantum field
theories on different branes must be compatible with the various
dualities.  One can use this observation to test the dualities.
Alternatively, assuming the various string dualities and the
consistency of the theory one can easily derive known results in
quantum field theory from a new perspective as well as many new
results.

\section{Conclusion}

During the last 30 years the structure of string theory has been
explored both in perturbation theory and non-perturbatively with
enormous success.  A beautiful and consistent picture has emerged.
The theory has also motivated many other developments, such as
supersymmetry, which are interesting in their own right.  Many of the
techniques that have been used to obtain exact solutions of field
theories were motivated by string theory.  Similarly, many
applications to mathematics have been discovered, mostly in the areas
of topology and geometry.  The rich structure and the many
applications are viewed by many people as indications that we are on
the right track.  However, the main reason to be interested in string
theory is that it is the only known candidate for a consistent quantum
theory of gravity.

There are two main open problems in string theory.  The first is to
understand the underlying conceptual principles of the theory -- the
analog of curved space-time and general covariance for gravity.  The
fact that we still do not understand the principles of the theory
makes this field different than others -- it is not yet a mature field
with a stable framework.  Instead, the properties of the theory are
being discovered with the hope that eventually they will lead to the
understanding of the principles and the framework.  The various
revolutions that the field has undergone in the last years have
completely changed our perspective on the theory.  It is likely that
there will be a few other revolutions and our perspective will change
again.  Indeed, fascinating connections to large N gauge theories are
currently being explored, which appear to be very promising. In any
case, the field is developing very rapidly and it is clear that an
article about string theory for the next centenary volume will look
quite different from this one.

The second problem, which is no less important, is that we would like
to make contact with experiment.  We need to find unambiguous
experimental confirmation of the theory.  Supersymmetry would be a
good start.

\section*{Acknowledgements}

We have benefited from many discussions with many colleagues
throughout the years.  Their insights and explanations were extremely
helpful in shaping our point of view on the subjects discussed here.
The work of JHS was supported in part by DOE grant
\#DE-FG03-92-ER40701 and that of NS by DOE grant \#DE-FG02-90ER40542.

\section*{References}

\begin{itemize}
\item
Selected books and reviews on supersymmetry:

J. Wess and J. Bagger, ``Supersymmetry and Supergravity,'' Princeton
Univ. Press, 1983 (Princeton series in Physics);

S. J. Gates, Jr., M. T. Grisaru, M.  Rocek, W. Siegel, ``Superspace:
or one thousand and one lessons in supersymmetry,'' Benjamin/Cummings,
1983 (Frontiers in Physics, Lecture Note Series 58);

P. West, ``Introduction to Supersymmetry and Supergravity,'' World
Scientific, 1986;

H. P. Nilles, ``Supersymmetry, Supergravity and Particle Physics,''
Phys.Rept. {\bf 110} (1984) 1;

H. E. Haber and G. L. Kane, ``The Search for Supersymmetry: Probing
Physics Beyond the Standard Model,'' Phys.Rept. {\bf 117} (1985) 75;

K. Intriligator and N. Seiberg, ``Lectures on Supersymmetric Gauge
Theories and Electric-Magnetic Duality'', hep-th/9509066,
{Nucl.Phys.Proc.Suppl. {\bf 45BC} (1996) 1};

M. E. Peskin, ``Duality in Supersymmetric Yang-Mills Theory'',
hep-th/9702094;

L. Alvarez-Gaume and F. Zamora, ``Duality in Quantum Field Theory and
String Theory,'' CERN-TH-97-257, hep-th/9709180.

\item
Selected books on string theory:

M. B. Green, J. H. Schwarz, and E. Witten, ``Superstring Theory,'' in
2 vols.  Cambridge Univ. Press, 1987 (Cambridge Monographs on
Mathematical Physics);

``Fields, Strings, and Duality'' (TASI 96), eds. C. Efthimiou and
B. Greene, World Scientific, 1997;

J. Polchinski, ``String Theory,'' in 2 vols.  Cambridge Univ. Press,
1998 (Cambridge Monographs on Mathematical Physics).

\end{itemize}

\end{document}